\documentclass[a4manuscript,12pt]{article}

\usepackage{bm}
\usepackage{bbm}
\usepackage[utf8]{inputenc}
\usepackage[english]{babel}
\usepackage{amssymb}
\usepackage{amsmath}
\usepackage{amsthm}
\usepackage{amsfonts}
\usepackage{mathrsfs}
\usepackage{dsfont}
\usepackage{hyperref}
\usepackage{commath}
\usepackage{graphicx}
\usepackage{text comp}
\usepackage{mathtools}
\usepackage{afterpage}
\usepackage{appendix}
\usepackage[linesnumbered,lined,boxed,commentsnumbered]{algorithm2e}

\usepackage{tabularx}
\usepackage{booktabs}
\usepackage[labelfont=bf,format=plain,justification=raggedright,singlelinecheck=false]{caption}
\theoremstyle{plain}
\newtheorem*{thm*}{Theorem}

\newtheorem*{lemma*}{Lemma}

\newtheorem*{corollary*}{Corollary}

\newtheorem*{prop*}{Proposition}

\newtheorem*{conjecture*}{Conjecture}

\newcommand{\R}{\mathbb{R}}
\newcommand{\RR}{\mathbb{R}}

\def\xvec{{\bf x}} 
\def\yvec{{\bf y}}

\def\gvec{{\bf g}}
\DeclarePairedDelimiter\floor{\lfloor}{\rfloor}

\usepackage{color}
\definecolor{mediumtealblue}{rgb}{0.0, 0.33, 0.71}

\definecolor{mypink1}{rgb}{0.858, 0.188, 0.478}

\everymath{\displaystyle}
\begin{document}

\title{Estimating Network Dimension When the Spectrum Struggles}

\author{
     Peter Grindrod \small{CBE}
\thanks{%
Mathematical Institute, 
University of Oxford, OX2 6GG, UK
     }
\and
        Desmond John Higham%
        %\texttt{email} \\
            \thanks{%
           School of Mathematics,
           University of Edinburgh,
           Edinburgh, EH9 3FD, UK
           %(\email{d.j.higham@ed.ac.uk}).
        }
        \and
        Henry-Louis de Kergorlay% 
    \thanks{%
           School of Mathematics,
           University of Edinburgh,
           Edinburgh, EH9 3FD, UK
           %(\email{hdekerg@ed.ac.uk})
           }
        }

\date{}

\maketitle
\begin{abstract}
  What is the dimension of a network?
  Here, we view it as the smallest dimension of 
  Euclidean space into which nodes can be embedded so that 
  pairwise distances accurately reflect the connectivity structure.
  We show that a recently proposed and extremely efficient algorithm
   for data clouds, 
  based on computing first and second nearest neighbour distances, can be 
  used as the basis of an approach for estimating the dimension of a network with weighted edges.
  We also show how the algorithm can be extended to unweighted networks 
  when combined with spectral embedding. 
  We illustrate the advantages of this technique over the widely-used  
  approach of characterising dimension by visually searching for a suitable gap in the spectrum
   of the Laplacian.
\end{abstract}

\section{Motivation} \label{sec:mot}
Given a network, it is often desirable to embed the nodes into Euclidean space 
so that distances between nodes reflect the connection strengths. 
Such an embedding may form a preprocessing step for visualization, clustering, or semi-supervised learning of labels 
\cite{BLSZ18}.
Spectral techniques, based on eigenvectors of a suitable Laplacian, are 
commonly used for the projection. In this case, choosing the 
dimension of the embedding, that is, the the number of eigenvectors used, 
is an important task. However, this task is
difficult to formalize, with a widely accepted 
rule of thumb being ``look for a gap in the spectrum'' \cite{spectClusteringTuto}.

In this work we investigate the use of the recently proposed two nearest neighbour, or \emph{twoNN}, algorithm in \cite{Facco17} as a means to inform the choice of dimension. 
The algorithm is designed to estimate the dimension of a cloud of data points, assuming the points are samples from 
a continuous manifold. It is extremely efficient, compared, for example, with box counting techniques, requiring only the pairwise distances between all first and second-nearest neighbours.
In the case of a weighted (undirected) network for which edge weights can be used to define the desired 
distances, the algorithm is directly applicable. We find that it performs well
on examples where a ground truth is available and where information from the Laplacian spectrum is at best ambiguous. 
In the case of unweighted networks, where edge weights are either present or absent, the algorithm can no longer be applied directly.
However, we show that 
on examples where a ground truth is available, useful estimates of 
the dimension can be recovered by spectrally embedding into successively higher dimensional Euclidean space and applying twoNN
at each stage. In cases where the ground truth has been contaminated by noise we find that, unlike the Laplacian spectrum, twoNN is robust and informative. 
In a final experiment on real data the twoNN estimate remains consistent when applied to 
a $K$ nearest neighbour  
binarization of  
the underlying weighted network.

\section{Set-up} \label{sec:setup}
Suppose we are given a non-negative $N\times N$  symmetric 
\emph{dissimilarity} matrix $M$.  We assume that  there are $N$ distinct underlying objects, $\xvec_i$, within some universal set $ S$, and that the element $M_{ij}$ for $i \neq j$ measures the dissimilarity between objects $\xvec_i$ and $\xvec_j$; so a larger $M_{ij}$ indicates that $\xvec_i$ and $\xvec_j$ are more dissimilar. 
Ideally elements of $M$ should also satisfy a triangle relationship; that is,  
$M_{ij}\le M_{ik}+M_{kj} $
for all distinct $i,j,k$.
This would hold by construction if elements of $M$ were derived from a metric space, $(S, \delta)$, containing the underlying $N$ objects; and $M_{ij}=\delta(\xvec_i, \xvec_j)$.
However, a triangle relationship is not necessary in what follows.

In some contexts, $M$ will be implied via  a weighted graph or a \emph{similarity} matrix: we have a
a symmetric $W \in \RR^{N \times N}$ 
such that each $W_{ij} \ge 0$ denotes the weight or strength of the connection between the  distinct objects $\xvec_i$ and  $\xvec_j$, with $W_{ii}=0$.  
In this case, a larger $W_{ij}$ indicates that $\xvec_i$ and $\xvec_j$ are more similar.

To convert between a dissimilarity matrix $M$ and a 
similarity matrix $W$, we may use, for example 
$M_{ij}=1/W_{ij}$, $M_{ij}=1-W_{ij}$,
or $W_{ij} = \exp(-M_{ij}\sigma^2)$ for some $\sigma$ \cite{KH22}.

Given 
a dissimilarity matrix, $M$, with no notion of $S$, we are often tasked with embedding $M$ into some Euclidean space: that is, we wish to define a set of real \emph{location} vectors, $\{ \yvec^{[i]} \}_{i=1}^{N}$ in $\RR^k$, such that the
Euclidean distances $\|\yvec^{[i]} -\yvec^{[j]}\|$ are monotonic functions of the $M_{ij}$.  
Ideally, the embedding dimension, $k$, should reflect the inherent dimension $d$ of the data.
There are many ways to embed such matrices, or equivalently the weighted graphs, and these usually involve some spectral analysis of a matrix related to $M$ or  $W$.

For unweighted graphs, where $W$ is binary and symmetric, there has been much interest within combinatorics. The dimension of a graph  is the smallest value of $d$ for which its vertices may be embedded in $\RR^d$ such  that the distances between the endpoints of each and every edge are equal to unity. Furthermore, in answer to a question of \cite{Erdos80},  any graph with less than ${d+2}\choose 2$ edges has dimension at most $d$; while the dimension of a graph with maximum degree $d$  is at most $d$ \cite{Frankl20}.

In this work we are concerned with the more practical question of how to compute a representative value for $d$ when 
it is likely that the network information is noisy or incomplete. 

\section{Spectral Embedding} \label{sec:spec}
Given the similarity matrix $W$,
it is common to embed the graph  
into a $k$-dimensional Euclidean space
using a spectral method. 
The embedding may be regarded as an $N \times k$ 
matrix 
\[
G= \left[\gvec^{[1]},\gvec^{[2]}, \ldots, \gvec^{[k]}\right],
\]
with orthonormal columns $\gvec^{[j]}\in \RR^N$.
The $i$th object $\xvec_i$ is given coordinates
according to the $i$th row of $G$; that is, 
\begin{equation}
\yvec^{[i]}
=
\left[
\begin{array}{c}
\gvec^{[1]}_i \\  \gvec^{[2]}_i \\ \vdots \\ \gvec^{[k]}_i
\end{array}
\right]
\in 
\RR^k.
\label{eq:yvecs}
\end{equation}
As described in \cite{belkin08}, 
it is natural to specify $G$ via 
\[
\arg \min_{\gvec^{[1]},\gvec^{[2]}, \ldots, \gvec^{[k]}}  \sum_{i=1,j=1}^m W_{ij} \|\gvec^{[i]}-\gvec^{[j]}\|^2 
\] 
with $G^T G = I$.
This leads to a solution where the columns $\gvec^{[i]}$ 
are given by eigenvectors corresponding to the $k$ lowest nonzero eigenvalues of the Laplacian
matrix $L=D-W$.
Here, $D \in \RR^{N \times N}$ is the diagonal matrix 
whose diagonal contains the row/column sums of $W$. 
We  note that  $L$ is self-adjoint and has a zero eigenvalue with geometric  multiplicity given by the number of connected components of the graph \cite{spectClusteringTuto}. Other related methods are available \cite{MKNS21}.   
This type of spectral embedding approach is closely related to principal component analysis and multi-dimensional scaling 
\cite{BG05,JC16}, and similar techniques are used for 
clustering \cite{CEH23,spectClusteringTuto},
ranking
\cite{Cucu21,KDMT22,Shur-preprint-flexible-pagerank},
subgraph detection \cite{BFMM022}
and graph visualization \cite{K05}.

How should we best choose the embedding dimension, $k$? Typically one might examine the spectrum of $L$, perhaps on a log scale, and search for an upward step, or \emph{gap}, discarding the eigenvalues/eigenvectors to the right
\cite{CG20,spectClusteringTuto}. In many circumstances, though, no clear step may be apparent. 

In essence, for a given choice of 
dimension $k$ spectral information from the Laplacian provides  
an optimal embedding, in a least squares sense.
But that spectral information does not always tell us how to  
choose the best $k$.

\begin{figure}[htp]
    \centering
    \includegraphics[width=0.495\textwidth]{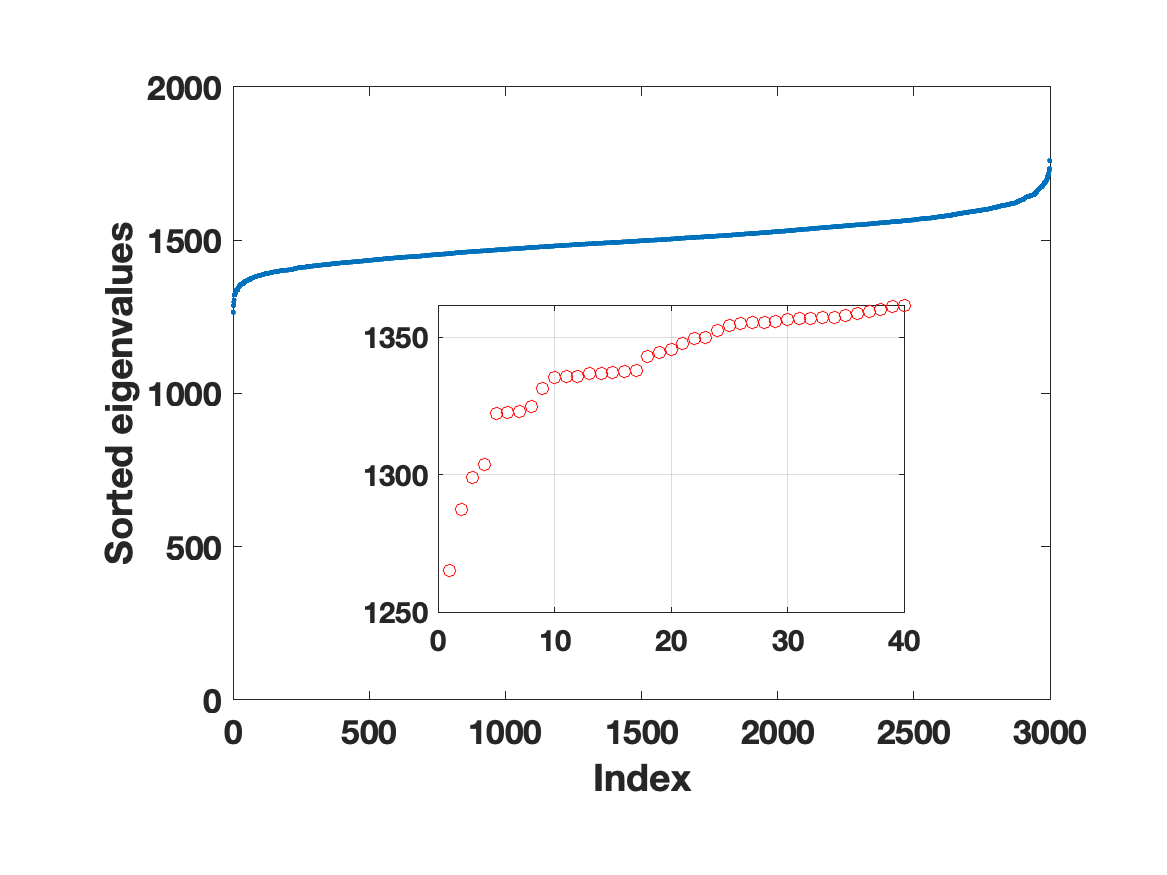}
     \includegraphics[width=0.495\textwidth]{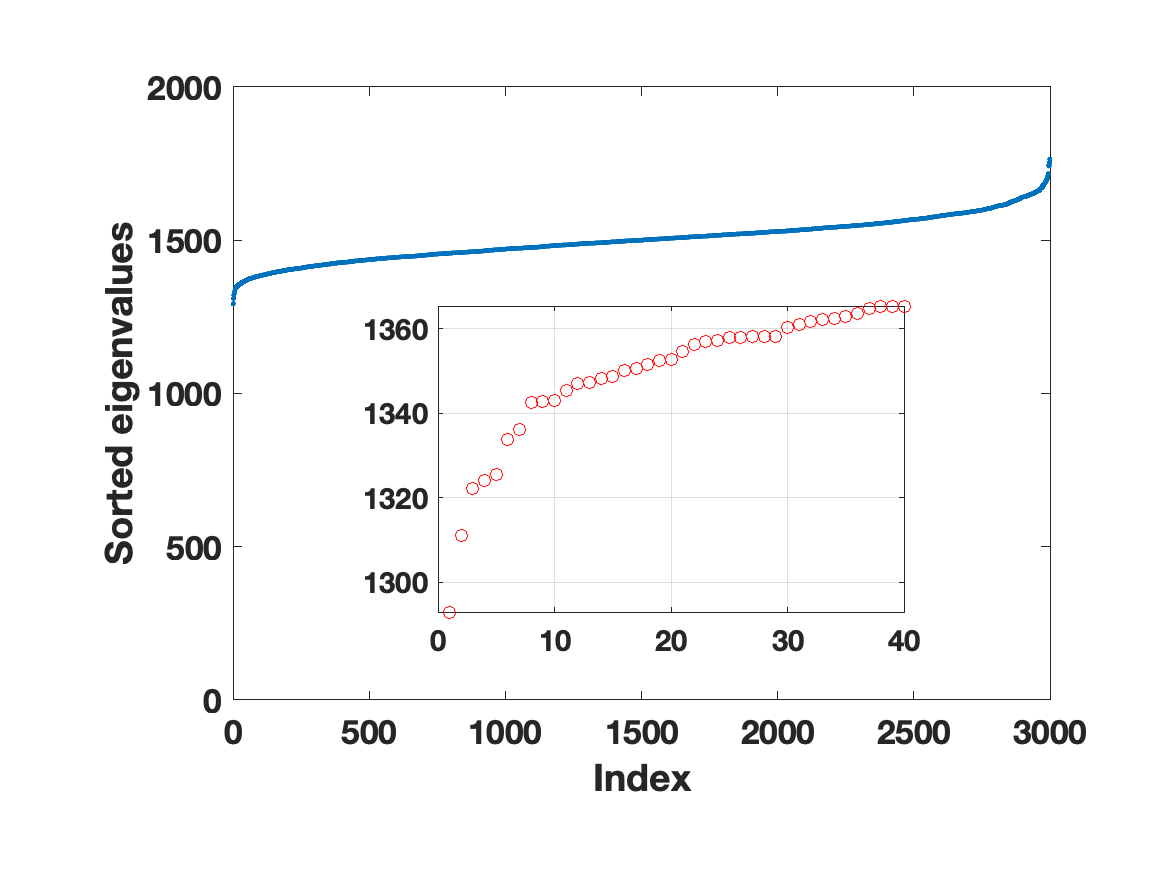}
    \caption{Ordered eigenvalues of the Laplacian based on inverse pairwise Euclidean distances between data clouds of $N = 3,000$ points in 
    $\RR^{25}$. Two independent instances are shown. 
    Interior plots focus on the first 40 nonzero eigenvalues.
    The spectrum gives no obvious argument for an embedding
    in dimension $25$.
    }
    \label{fig:Lap}
\end{figure}

In Figure~\ref{fig:Lap} we show the Laplacian spectrum for two examples where $N=3,000$. Here we 
sampled vectors $\xvec_i \in \RR^{25}$ with components chosen independently and uniformly at random in $[0,1]$, and set $W_{ij}=1/ \| \xvec_i-\xvec_j \|$, for $i\not= j$.  So, by design, we hope that the graph (or the matrix) will be embeddable in $\RR^{25}$.  Yet this is not reflected by the spectrum shown in Figure~\ref{fig:Lap}: an ``eyeball'' search for a gap in the spectrum would not highlight a dimension of $25$.

\section{TwoNN and Weighted Graphs}\label{sec:weight}

In this work we investigate the approach of choosing the embedding dimension with the twoNN method, which was developed 
in \cite{Facco17}
to estimate what those authors refer to as the intrinsic (fractal) dimension of sparse point clouds in very high dimensional Euclidean spaces;
this may be viewed as the dimension of an underlying continuous manifold from which the data points are sampled.
We refer to \cite{BLGS21} for further discussion of intrinsic dimension and connections to learning theory and  
topological data analysis. 
With twoNN, 
for all objects in the point cloud one finds the distances to the nearest neighbour and to the second nearest  neighbour. The ratio, $\mu >1$,  of the latter distance to the former, calculated separately for all points, produces an empirical cumulative distribution, say 
$F^{\mathrm{emp}}(\mu)$. This can be compared with the expression $F(\mu) = (1 - \mu^{-d})$ that the authors derived under the assumption of 
local uniform sampling density from a space of dimension $d$. Comparing the empirical and exact cumulative distributions allows us to estimate the dimension $d$.
If we let $\mu_i$ denote the second-to-first nearest neighbour distance ratio for node $i$, 
and let $\sigma$ denote a permutation vector such that $\mu_{\sigma(i)}$ are in ascending order, then 
the empirical cumulative distribution 
has $F^{\mathrm{emp}}(\mu_{\sigma(i)})  = i/N$, and the points
$[ \log \mu_{\sigma(i)}, - \log( 1 - F^{\mathrm{emp}}(\mu_{\sigma(i)}) )]$
should lie on a line of slope $d$. 
So the points
\begin{equation}
d_i = -\frac{\log(1 - i/N)}{\log \mu_{\sigma(i)} }
\label{eq:didef}
\end{equation}
are estimates for the dimension $d$.

\begin{figure}[htp]
    \centering
    \includegraphics[width=0.495\textwidth]{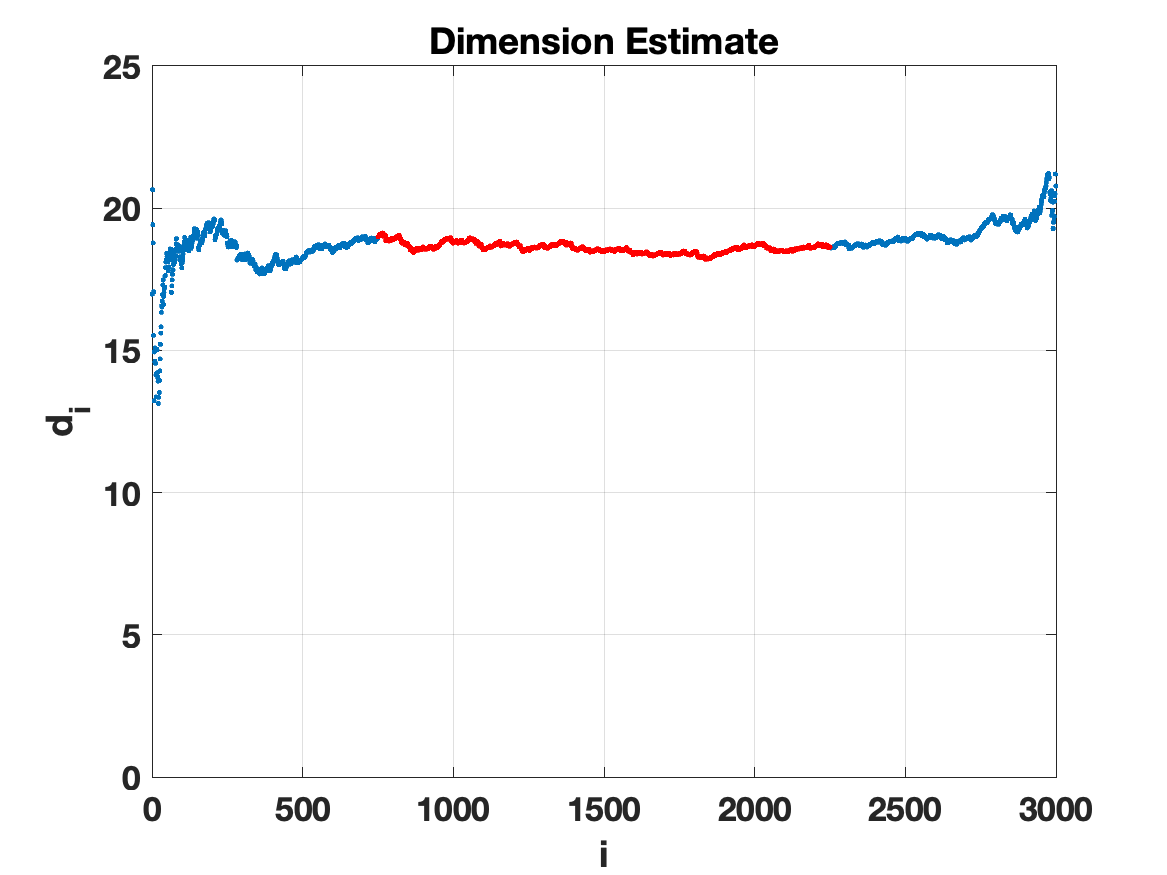}
    \includegraphics[width=0.495\textwidth]{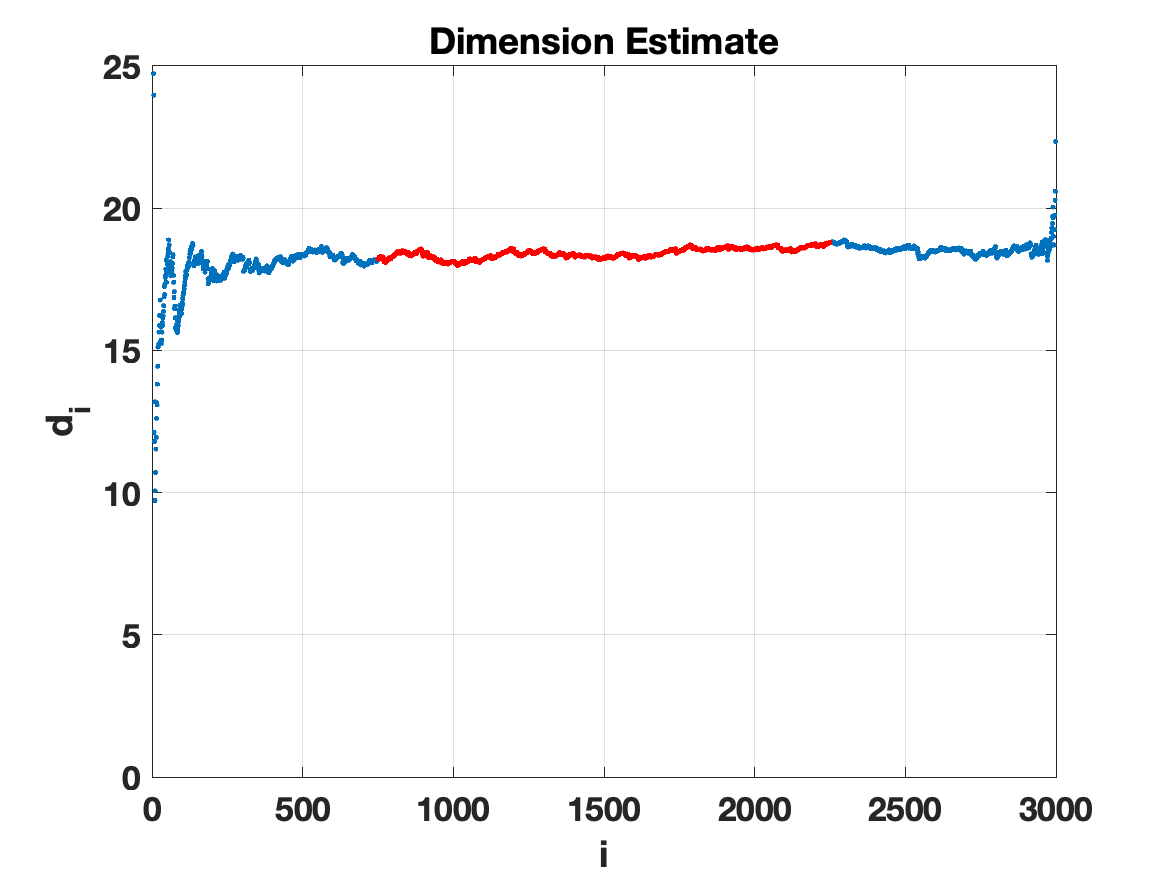}
    \caption{TwoNN estimates for the embedding dimension. We show dimension estimates 
    $d_i$ in (\ref{eq:didef}) versus the index $i$. 
    }
    \label{fig:twonn}
\end{figure}

When we are presented with data in the form of a dissimilarity matrix $M$,
we may infer pairwise distances directly: the two nearest neighbour distances of an object simply correspond to the two smallest elements of the corresponding row of $M$.
In Figure~\ref{fig:twonn}
we illustrate the method on the two data sets used in Figure~\ref{fig:Lap}. Here, we set 
$M_{ij} = 1/W_{ij} = \| \xvec_i-\xvec_j \|$.
The plots show $d_i$ in (\ref{eq:didef}) against $i$.
As discussed in 
\cite{Facco17}, 
for small $i$ we expect sampling errors to dominate, with too little information in play; whereas if $i$ is too large then we would see the lack of very large-scale distance differences affect the estimate (since the distances are  globally bounded). 
So a sweetspot is desirable, where the estimate for $d$ is relatively static. In Figure~\ref{fig:twonn} we 
highlight in red the portion of the curve where $N/4 \le i \le 3 N/4 $, which leads to stable estimates, and throughout this work we use
the mean of $d_i$ over this range as our overall estimate of $d$.
For Figure~\ref{fig:twonn} we obtain estimates of $18.6$ and $18.4$.
This approach appears much more definitive than a visual search for a spectral \emph{gap}.

We emphasize that in Figures~\ref{fig:Lap} and \ref{fig:twonn}, 
the $N=3,000$ data points $\xvec_i$ were  
generated within the cube $[0,1]^{25}$. Yet $2^{25}=33,554,432$, so this set of points will hardly get close to any of the extremities.
Hence, this example is challenging---high dimensional spaces are very lonely places.
Moreover, $25$ should be considered as a hard upper bound for any estimate of $d$, with the distance/dissimilarity matrix likely to be more consistent with a smaller dimension. 

\begin{figure}[htp]
    \centering
    \includegraphics[width=0.495\textwidth]{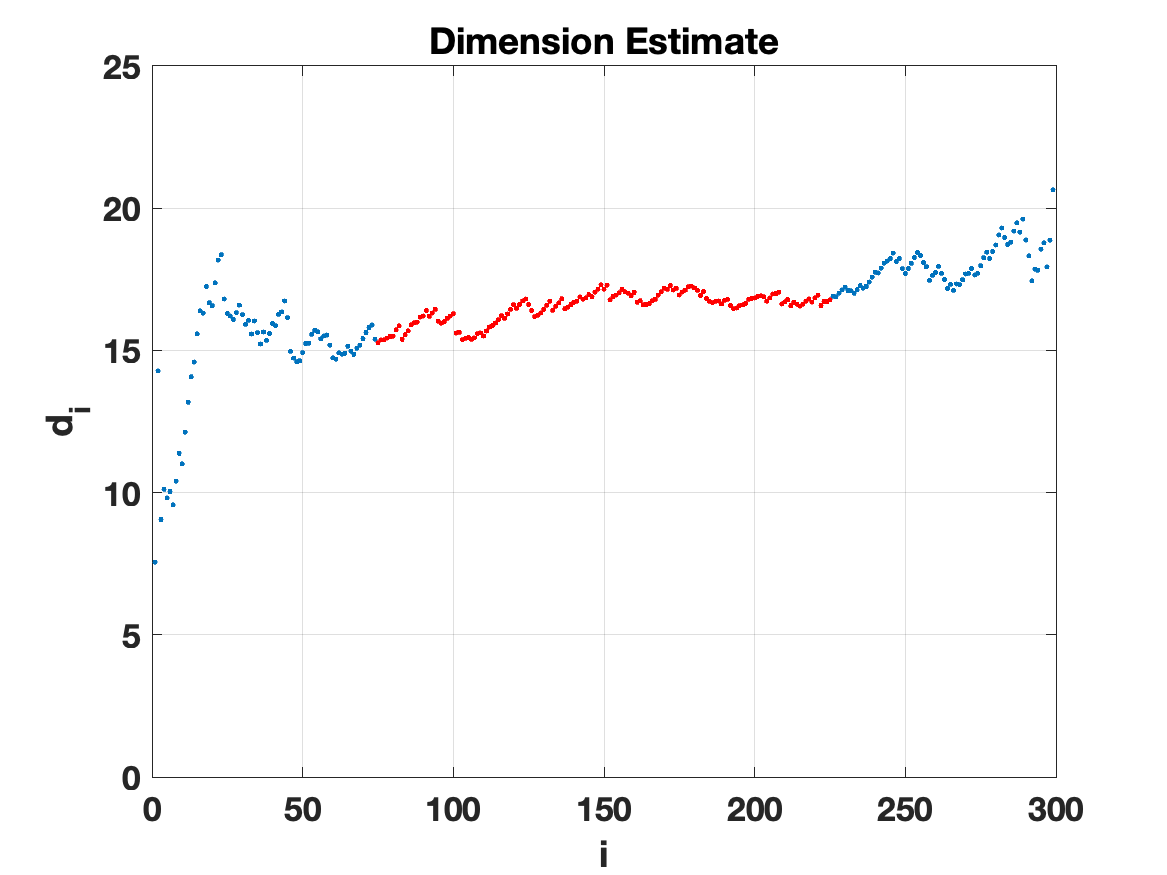}
    \includegraphics[width=0.495\textwidth]{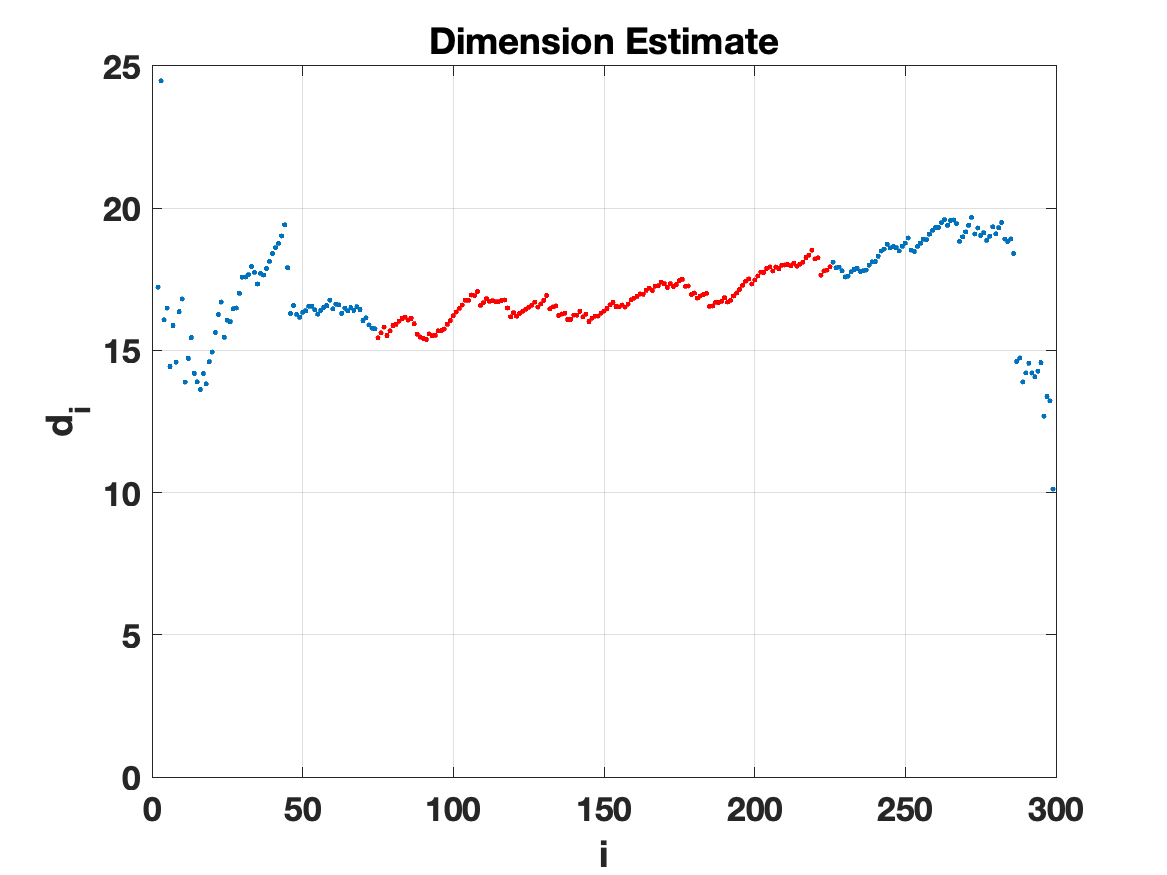}
    \caption{As for Figure~\ref{fig:twonn}, but with the number of data points reduced from $N = 3000$ to $N = 300$.
    }
    \label{fig:twonn300}
\end{figure}

From a theoretical perspective, the following conditions are known to be sufficient for
 a discrete graph Laplacian to converge spectrally to the Laplace-Beltrami operator over an underlying sampling manifold, e.g. \cite{Trillos18SpectClustering}:
\begin{align}\label{kernel conditions for spect convergence of Laplacian}
\begin{cases}
\pmb{\eta}(0)>0 \text{ and } \pmb{\eta} \text{ is continuous at }0,\\
 \pmb{\eta}\text{ is non-increasing},\\
 \int_{\R^d}\eta(x)\abs{\langle x,e_1\rangle }^2dx<\infty,
\end{cases}
\end{align}
where $\eta$ is the radially symmetric kernel function such that $W_{ij}=:\eta(x_i-x_j)$ for all $i,j\in \{0,\dots,N\}$, and where $\pmb{\eta}$ is the radial profile, or the shape, of $\eta$, i.e., $\eta(x)=\pmb{\eta}(||x||)$ for all $x\in \R^d$. 
However, in a typical practical setting, such conditions cannot be validated, in which case there is 
no guarantee that the spectral embedding is consistent. 
We note that the estimate of $d \approx 18.5$ from Figure~\ref{fig:twonn} provided by twoNN appears acceptable, even though the conditions in (\ref{kernel conditions for spect convergence of Laplacian}) for $W$ are not satisfied. 
Moreover, we show in Figure~\ref{fig:twonn300} an experiment where $N$ is reduced to $300$. Here, twoNN still produces reasonable estimates of 
$16.5$ and $16.8$. 

We also note that the algorithm is insensitive to smooth rescaling of the distance measure, in the sense that 
if $r_2$ and $r_1$ are two small distances then
\[
\frac{f(r_2)}
{f{(r_1)}}
\approx 
\frac{f(0) + r_2 f'(0)}
{f(0) + r_1 f'(0)}
= 
\frac{r_2}
{r_1},
\]
for any differentiable function $f$ such that $f(0) = 0$ and $f'(0) \neq 0$.

We therefore suggest that direct application of the twoNN method from \cite{Facco17} on a weighted network provides reliable
information 
for choosing a spectral embedding dimension $k$, and at the very least may be  regarded as a back-up procedure 
or sanity check for the widely-adopted approach of visually inspecting the spectrum of the Laplacian. 
In the remainder of the manuscript we focus on the more challenging case of an unweighted network, where the algorithm is not directly applicable.

\section{TwoNN and Unweighted Graphs}\label{sec:unweight}

With an unweighted graph, it remains natural to seek a node embedding  
such that nearby nodes are connected and distant nodes are unconnected.
For example, we may postulate that the connectivity structure in the graph arose from some sort of (unobserved) binarization  
mechanism, such as $K$ nearest neighbour or radius-based thresholding (geometric).
For an unweighted graph, the notion of first and second nearest neighbour  
is not immediately applicable. 
Hence, we advocate an indirect approach where spectral embedding is used as an intermediate step.
Here, we look for the largest $s$ such that after spectrally embedding into dimension $s$ 
the twoNN algorithm also delivers an estimate close to $s$ for the dimension.
Intuitively, if we spectrally embed into a dimension that is unnecessarily small, then
twoNN will reproduce this dimension, whereas if we 
spectrally embed into a dimension that is unnecessarily large, then
twoNN will find the appropriate, smaller dimension.

To be concrete, in Algorithm~\ref{alg:spectwoNN} we outline the
steps involved in spectral embedding followed by the application of twoNN.
Our overall approach is then to apply Algorithm~\ref{alg:spectwoNN} for $s = 2,3,\ldots$
and observe how the twoNN estimate $d^\star$ compares with the embedding dimension $s$, stopping when 
$d^\star$ plateaus as a function of $s$. 

\begin{algorithm}

\SetKwInOut{Input}{input}\SetKwInOut{Output}{output}

\Input{Similarity (adjacency) \textbf{unweighted} matrix, $W \in \RR^{N \times N}$; trial embedding dimension, $s$}
\BlankLine

\Output{twoNN dimension estimate, $d^\star$}
\BlankLine

  $\Delta \leftarrow$ graph Laplacian obtained from $W$\;
  $\gvec^{[1]}, \dots, \gvec^{[k]}\leftarrow$ $k$ eigenvectors associated to the $k$ first non-trivial eigenvalues of $\Delta$\;
  $\yvec^{[1]}, \dots, \yvec^{[N]}\leftarrow$ $N$ embedding locations from (\ref{eq:yvecs})\;
  \For{$i=1:N$}{
  $r^1_i,r^2_i \leftarrow$ $1$st and $2$nd nearest neighbour distances for $\yvec^{[i]}$\;
  $\mu_i \leftarrow \frac{r^2_i}{r^1_i}$\;
  }
  $\sigma\leftarrow$ permutation such that $\mu\circ\sigma$ is ordered in ascending order\;
  $d_1, d_2, \ldots, d_N \leftarrow$ dimension estimates from (\ref{eq:didef})\;
  $d^\star \leftarrow$ mean of $d_i$ over $N/4 \le i \le 3N/4$\;
 \caption{Combination of spectral embedding and twoNN that can be used iteratively to estimate the intrinsic dimension of an unweighted network.\label{alg:spectwoNN}}
\end{algorithm}

In the next subsection, we test this approach as follows. 
First, we create a ground truth by 
starting with a node sampling.
We then 
binarize the pairwise distance information
using a $K$ nearest neighbour construction.
We consider two different sampling settings; 
first from the standard normal distribution on $\R^d$ and second using components that are uniform on $[0,1]$.

The case of binarization via a 
geometric graph construction is considered in subsection~\ref{subsec:geom}.

\subsection{Results for $K$ Nearest Neighbour Construction}\label{subsec:KNN}

Given $K$, let $A \in \RR^{N \times N}$ be the adjacency matrix of a $K$ nearest neighbour ($K$NN) graph constructed as follows.
Associate $\xvec^{[i]}$ with a vector in $\RR^d$ with independently chosen entries and 
record an edge between every distinct pair of nodes if one of them 
is among the $K$ nearest neighbours of the other.
There exist alternative constructions for building a $K$ nearest neighbour graph, see, for example, \cite{kNNOptimalConstruction}, all yielding a symmetric affinity matrix, and we expect the computational results and choices of parameters to be equivalent for these alternative constructions.

It is known, \cite{kNNNumberOfNeighboursNeeded,kNNConnectivity},  that unless $K=\Omega (\log N)$, the $K$ nearest neighbour graph (any construction) is almost surely not connected for $N$ sufficiently large. It is also known, \cite{spectConvergencekNNLaplacian}, that one must choose $K=\omega(\log N)$ in order for the discrete graph Laplacian to converge spectrally to the underlying continuous Laplace-Beltrami operator, whose spectrum characterizes the geometry of the underlying sampling domain. If this condition is not satisfied, the eigenvectors of the discrete Laplacian are not guaranteed to converge to the eigenfunctions of the associated Laplace-Beltrami operator; hence the spectral embedding in (\ref{eq:yvecs}) is not guaranteed to be accurate. Based on this observation, it makes sense to choose $K=\omega(\log N)$. In our experiments, we fix
$d = 25$, 
$N=3,000$, and choose $K:=\floor{30 \log N}$.

\begin{figure}[htp]
    \centering
    \includegraphics[width=1\textwidth]{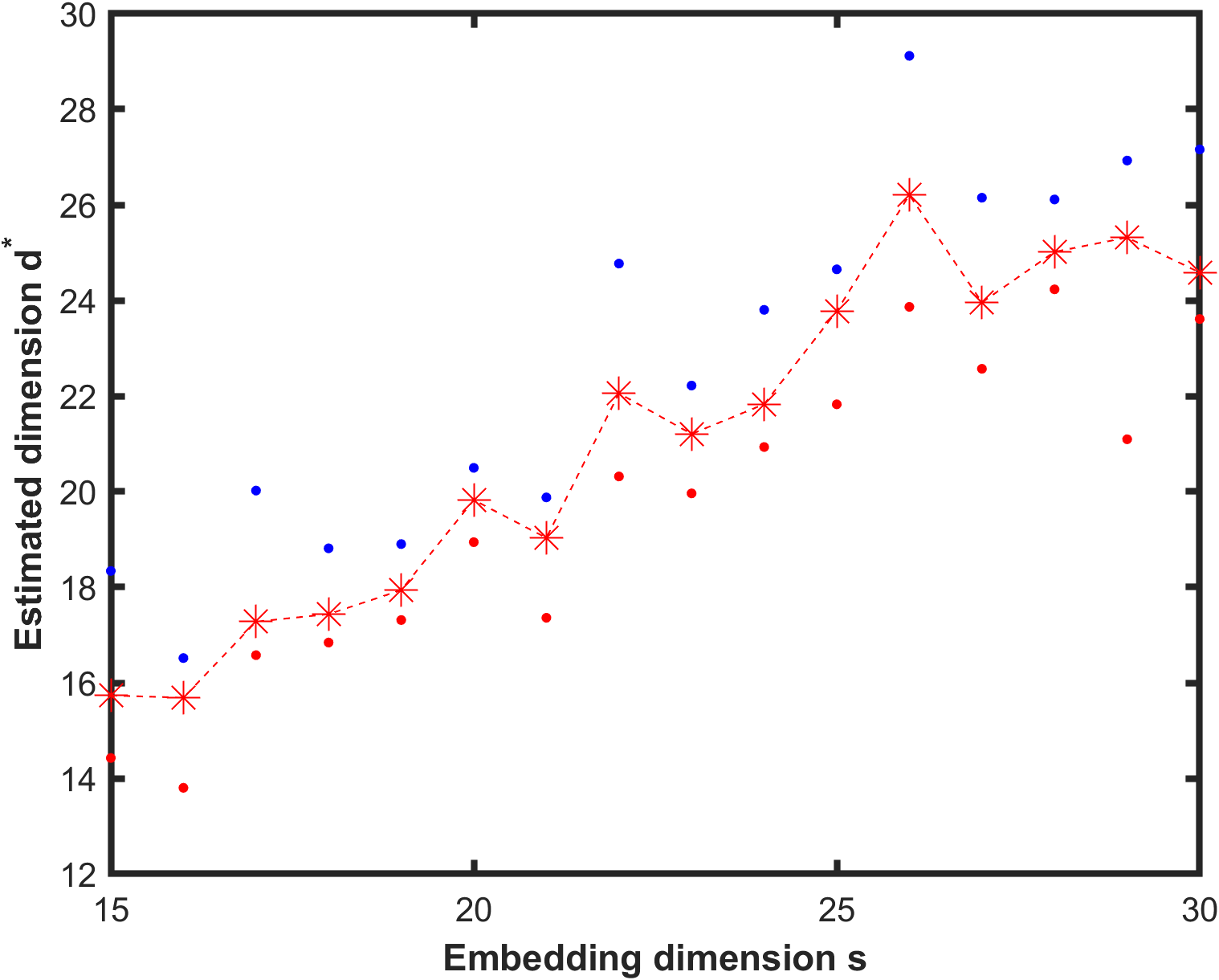}
    \caption{Results from Algorithm~\ref{alg:spectwoNN} with dimension $s$ varying from  
    $15$ to $30$ on the horizontal axis. Here we sampled $N=3000$ points from a Gaussian in $\R^{25}$ and used a $K$NN construction to produce an unweighted graph.
    %Each experiment is repeated \textcolor{red}{????} times. 
    The red  dots, blue dots and asterisks show the minimum, maximum and mean $d_i$ value for each $s$.
    }
    \label{fig:alg1KNNgauss}
\end{figure}

\begin{figure}[htp]
    \centering
    \includegraphics[width=1\textwidth]{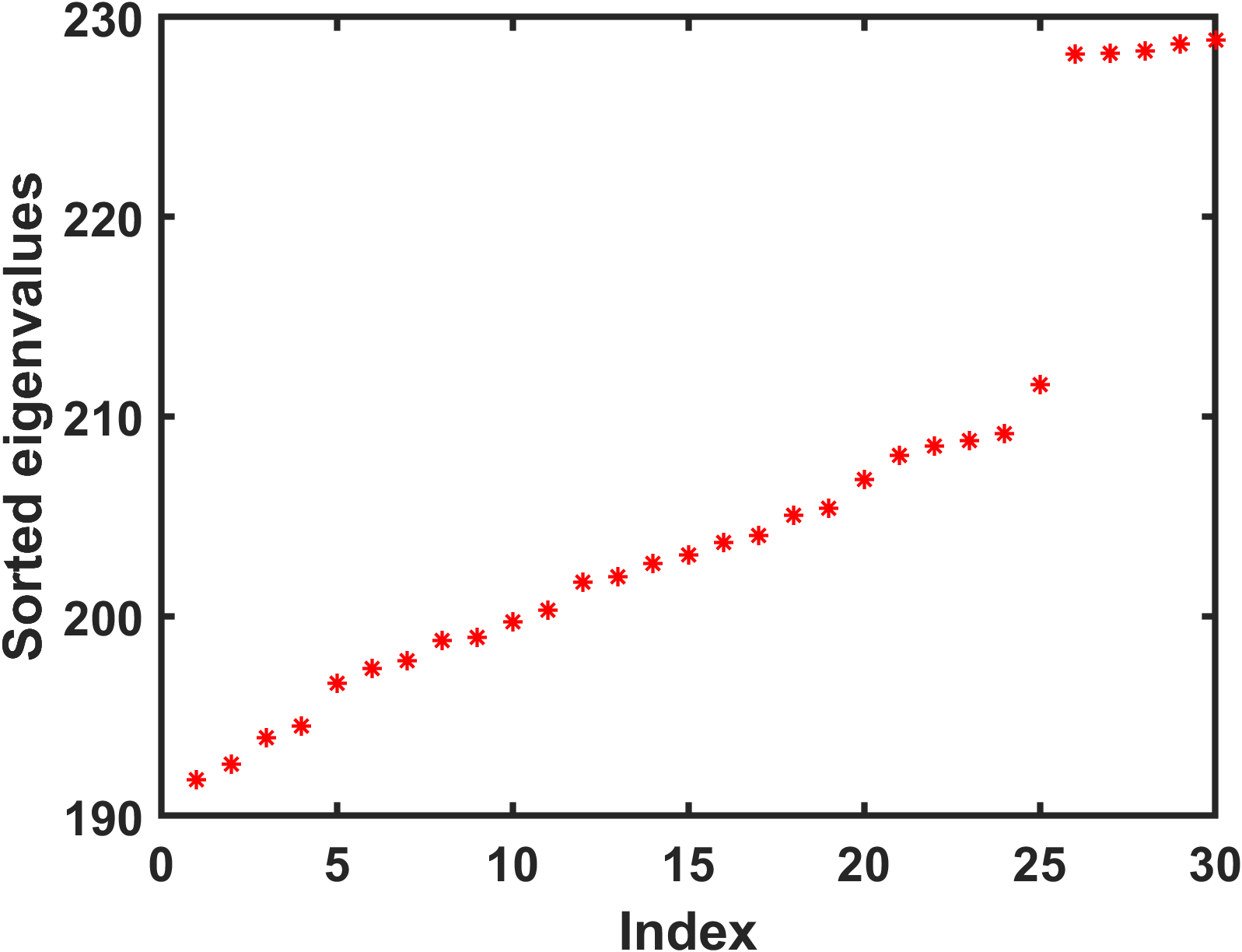}
    \caption{First thirty ordered nonzero eigenvalues for the Laplacian of an unweighted $K$NN graph from the experiment in 
    Figure~\ref{fig:alg1KNNgauss}, showing a spectral jump at dimension $d=25$.
    }
    \label{fig:KNNgapgauss}
\end{figure}

Figure~\ref{fig:alg1KNNgauss} shows the estimation of the intrinsic dimension from Algorithm~\ref{alg:spectwoNN}, for each embedding dimension in $\{15,16,\dots,30\}$, in the case where we sample $N=3000$ points from a Gaussian in $\R^{25}$. We see that the slope increases and stabilises around $25$. Figure~\ref{fig:KNNgapgauss} shows that the corresponding graph Laplacian 
has a spectral jump at $d=25$. 
In the next section we will show that this spectral information degrades in the presence of noise.

\begin{figure}[htp]
    \centering
    \includegraphics[width=1\textwidth]{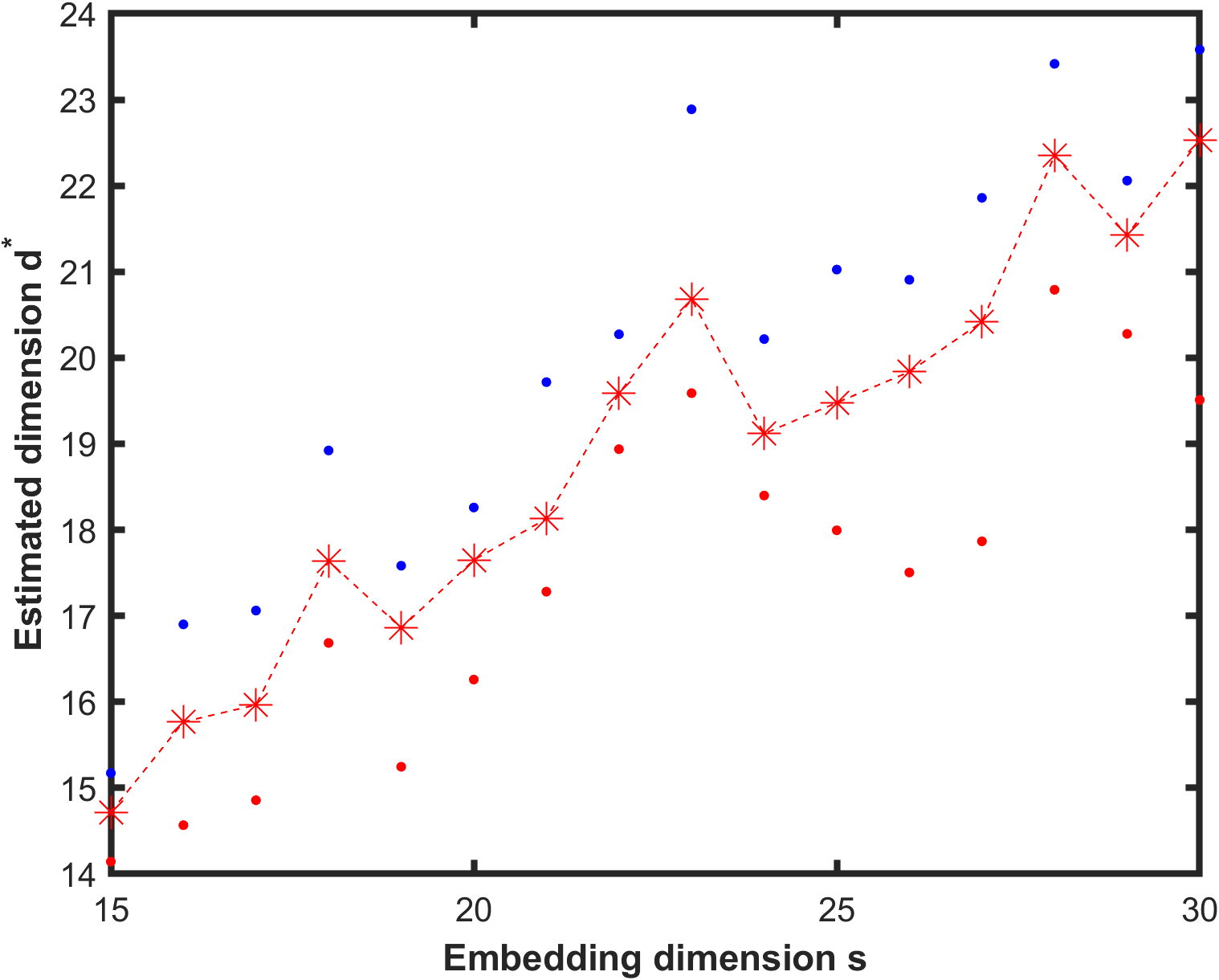}
    \caption{As for Figure~\ref{fig:alg1KNNgauss} with components sampled uniformly in $[0,1]$.
    }
    \label{fig:alg1KNNuni}
\end{figure}

\begin{figure}[htp]
    \centering
    \includegraphics[width=1\textwidth]{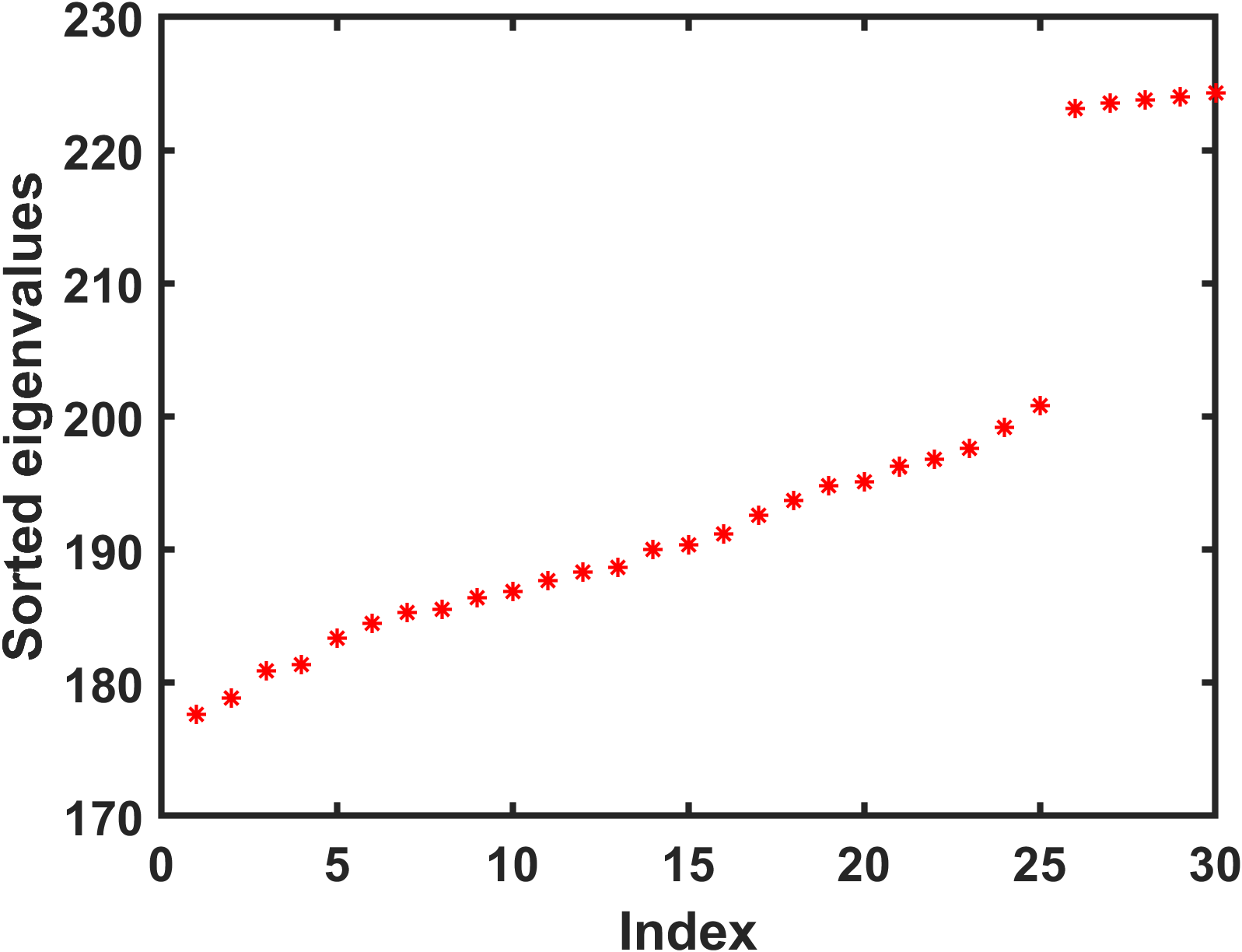}
    \caption{As for Figure~\ref{fig:KNNgapgauss} with components sampled uniformly in $[0,1]$.
    }
    \label{fig:KNNgapuni}
\end{figure}

We obtain similar pictures in the case where we sample the data point components uniformly in $[0,1]$, as indicated in 
Figures~\ref{fig:alg1KNNuni} and \ref{fig:KNNgapuni}.

\subsection{Geometric Graph Constructions and the Curse of Dimensionality}\label{subsec:geom}

For random geometric graphs where edges are inserted when the pairwise distance is below a radius $r>0$, 
there are well-known results on connectivity; see, for example, \cite{randGraphsConnectivity,randGraphs}, which provide asymptotic conditions.
However, choosing $r$ suitably in practice remains challenging, as we are typically in a non-asymptotic regime, and becomes unfeasible as the intrinsic dimension of the data becomes large. This is an example of what is commonly referred to as the curse of dimensionality. We illustrate the effect with the following geometric construction, which helps to explain what 
may go wrong when the dimension is large.

Given the unit cube $C$ in $\R^d$, consider the $d$-dimensional rectangle $R$ given by $[0,1]\times [r,1-r]^{d-1}$. Let $e_1,\dots,e_d$ be orthonormal vectors in the $d$ directions spanned by the axes of $\R^d$. The rectangle $R$ has exactly two faces per direction ($(d-1)$-dimensional faces), and we can assume, without loss of generality, that the interval $[0,1]$ of $R$ is along the $e_1$ direction, so that the two faces in the $e_1$ direction are squares contained in two faces of the unit cube. These faces can be written as
$$
\{(\gamma_1,\dots,\gamma_d)\ |\ \gamma_1=\epsilon,\ \forall\ k\in\{2,\dots,d\},\ \gamma_k\in [r,1-r]\},
$$
where $\epsilon\in \{0,1\}$.
Letting $k\in \{2,\dots,d\}$, the two faces associated to the $e_k$ direction can be written as
$$
\{(\gamma_1,\dots,\gamma_d)\ |\ \gamma_1\in [0,1],\ \gamma_k=\gamma,\ \forall\ j\not\in \{1,k\},\ \gamma_j\in [r,1-r]\},
$$
where $\gamma\in \{r,1-r\}$ is the constant value of the $k$th coordinate of every element of a given face in the $e_k$ direction.

For a direction $k\in \{2,\dots,d\}$ and a face in the $e_k$ direction given by the choice of $\gamma\in \{r,1-r\}$, define the $d$-dimensional rectangle
$$
R_{k,\gamma}:=\{(\gamma_1,\dots\gamma_d)\ |\ \gamma_1\in [0,1],\ \gamma_k\in I_{\gamma},\ \forall\ j\not\in\{1,k\},\ \gamma_j\in [1/3,2/3]\},
$$
where
$$
I_{\gamma}:=\begin{cases}
[1-r,1], \text{ if }\gamma=1-r,\\
[0,r], \text{ if }\gamma = r.
\end{cases}
$$
Each such rectangle ``closes the gap between two adjacent corners of the unit cube", such that the set obtained, after removing from the unit cube $R$ and every $R_{k,\gamma}$, $k\in \{2,\dots,d\}$, $\gamma\in \{r,1-r\}$, consists of $2^{d-1}$ connected components.
Furthermore, the volume of the union of these sets is
$$
V(R\cup \cup_{k,\gamma}R_{k,\gamma})=2(d-1)(1/3)^{d-2}r+(1-2r)^{d-1},
$$
which gets very small very quickly as the dimension $d$ increases, even for small values of $r$. For instance, for $d=10$ and $r=0.1$, this volume is $0.135$, for $d=15$, it is $0.0440$, and for $d=25$, it is $0.0047$.

 Thus, unless the dimension of the data set is small, most of the sampled points will be equi-distributed among the $2^{d-1}$ connected components of the unit cube after having removed $R\cup\cup_{k,\gamma}R_{k,\gamma}$, which as a union do not form a connected set. It is not possible to create a connected geometric graph from such a sample, unless we choose $r>0$ trivially to be so large that its size is comparable to the side width of the unit cube, in which case, the generated graph will be close to being complete. This issue does not arise with the $K$NN construction, since the number of neighbours is fixed and determined by the choice of $K$, which allows us to choose $K$ such that the graph is both connected and sufficiently sparse, even in high dimensions.

This thought experiment also indicates that, in our context, it is 
unreasonable to assume that a high-dimensional unweighted graph arose from a random geometric graph 
construction---sparsity and connectivity are unlikely to hold simultaneously.
A $K$NN construction is thus a more realistic binarization mechanism.

\subsection{Observations on non-uniform sampling densities and $K$NN constructions}

Neither the uniform distribution on the unit cube nor the Gaussian distribution on $\R^d$ are uniform in the underlying metric space (here $\RR^d$ with the Euclidean distance in both cases). Hence the sampling density will not be approximately constant in the neighbourhood of every sampled point, a condition that was assumed in the derivation of the twoNN algorithm in \cite{Facco17}. In the case of the uniform distribution on the unit cube, the sampling density fails to be approximately constant around points close to the boundary, while in the case of a Gaussian distribution, the sampling density is never approximately constant, in any neighbourhood.
However, the $K$NN graph construction allows us to cancel to some extent the negative effect of the non-uniformity of the sampling densities, as observed by the accurate estimations of the intrinsic dimension of the data in Section~\ref{subsec:KNN}. By analogy with a geometric graph, where points are connected if they are at a distance less than a fixed bandwidth parameter $r>0$, a $K$NN graph can be thought of as connecting points if they are at a distance less than a varying bandwidth parameter, inversely proportional to the values taken by the density in a given neighbourhood of those points. Every point is connected to roughly the same number of neighbours, mimicking a geometric graph if the points had been sampled from a uniform density, canceling the presence of irregularities on the sampling domain, such as boundaries. We can thus expect the spectrally embedded points in twoNN to be approximately uniformly distributed in the case of a $K$NN construction, even if the points were sampled from a non-uniform density on the original domain.

\begin{figure}[htp]
    \centering
    \includegraphics[width=1\textwidth]{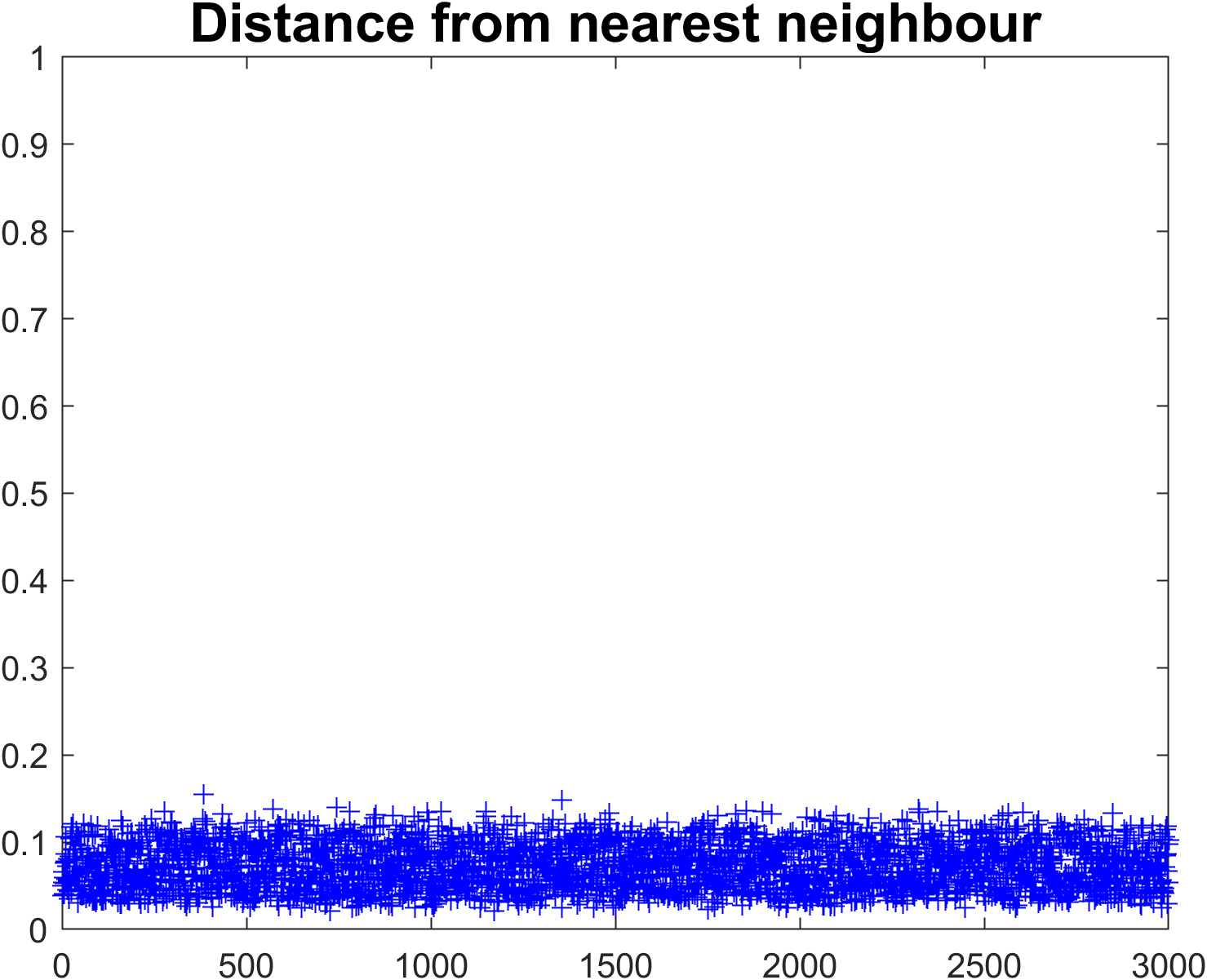}
    \caption{Nearest neighbour distance for each of the $3000$ points uniformly sampled from the unit cube in $\R^{25}$, spectrally embedded in $\R^{30}$, obtained from a $K$NN construction. We observe that the nearest neighbour distances concentrate tightly around an expected distance. This suggests the embedded points are close to being uniformly distributed, in which case their nearest neighbour distances would follow an approximately normal distribution.
    }
    \label{fig: nn distances}
\end{figure}

\begin{figure}[htp]
    \centering
    \includegraphics[width=1\textwidth]{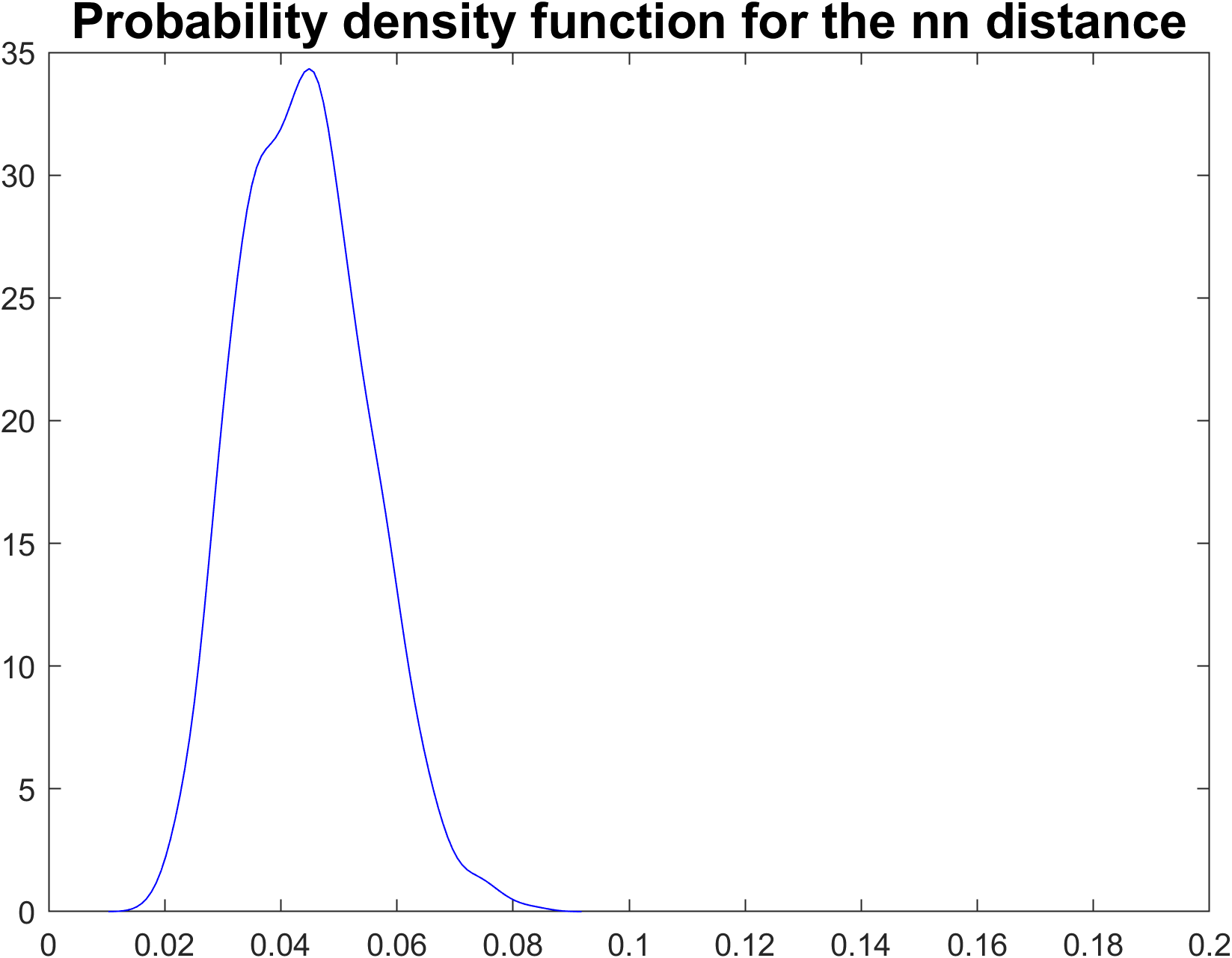}
    \caption{ 
    Sampling density of the nearest distance neighbours from the spectral embedding of dimension $30$, from points sampled uniformly from the unit cube in $\R^{25}$, obtained from a $K$NN construction as in Figure~\ref{fig: nn distances}. The horizontal axis indicates the value of the nearest neighbour distance, and the vertical axis indicates the number of points whose nearest neighbour distance satisfies this value. The normal-like distribution suggests that the points in the spectral embedding are approximately uniformly distributed.
    }
    \label{fig: probability density nn distance}
\end{figure}

We test this hypothesis by looking at the distribution of the embedded points $\{\yvec^{[i]}\ |\ i\in \{1,\dots,n\}\}$ in $\R^s$, $s>d$, in the case where the points are uniformly sampled from the unit cube in $[0,1]^{d}$. In the figures below, we choose $d=25$ and $s=30$.
%A lack of uniform distribution for the points in the image space would reflect the sampling density not being constant around the points near the boundary in the sampling domain, which does not happen with a $K$NN construction.
If the sampling density is uniform, then the nearest neighbour distances of the sampled points should concentrate highly around their expected value, forming an inverse exponential distribution centered around the expected nearest distance. In Figure~\ref{fig: nn distances}, we plot and compare the nearest distances for all embedded points. We observe that the nearest neighbour distances are indeed highly concentrated around their expected value, which does suggest that the embedded points in the image space must be approximately uniformly distributed. In Figure~\ref{fig: probability density nn distance}, the approximate uniform distribution of the embedded points is confirmed by the observation that the the nearest neighbour distances concentrate around their expected value following a Gaussian-like distribution.

\section{Tests on Noisy Data}\label{sec:noisy}

The experiments in section~\ref{sec:unweight}  assume that the similarity matrix is recorded with perfect information. In that idealized setting, 
we found that the twoNN algorithm did not provide a better method to infer the intrinsic dimension of the data than a simple spectral gap reading. In this section we test a more realistic scenario where the information is recorded with noise.

For every entry of the upper triangular part of the binary adjacency matrix $A$, we change independently the value from $0$ to $1$ or from $1$ to $0$ with probability $p$, and otherwise leave it unchanged. To keep the matrix $A$ symmetric,  
the upper triangular part of $A$ determines the remaining entries of $A$. 
In other words, each edge and each missing edge is flipped independently with probability $p$.
In our experiment, we used intrinsic dimension $d=15$ and sampled $N=17000$ vectors uniformly at random from the unit cube $[0,1]^d$. We generated $A$ from a $K$NN construction, picking $K=\floor{30\log(N)}$ as before, and introduced noise to $A$ as described above with parameter $p=0.01$.

\begin{figure}[htp]
    \centering
    \includegraphics[width=1\textwidth]{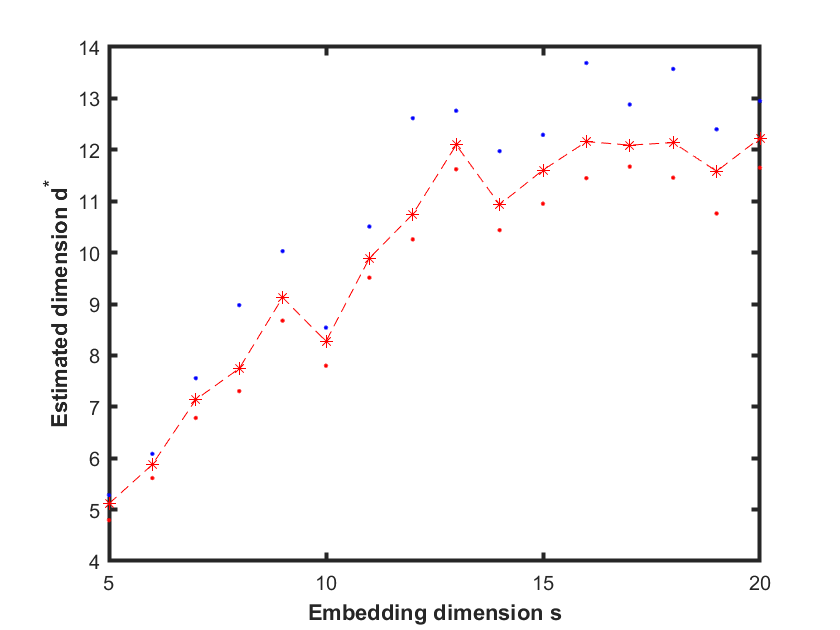}
    \caption{
    Results from Algorithm~\ref{alg:spectwoNN} with dimension $s$ varying from  
    $5$ to $30$ on the horizontal axis. Here we sampled $N=17000$ points from the unit cube $[0,1]^{15}$, used a $K$NN construction to produce an unweighted graph and then flipped each edge/missing edge with independent probability $p = 0.01$.
    %Each experiment is repeated \textcolor{red}{????} times. 
    The red  dots, blue dots and asterisks show the minimum, maximum and mean $d_i$ value for each $s$.
    }
    \label{fig:2NNnoise}
\end{figure}

\begin{figure}[htp]
    \centering
    \includegraphics[width=1\textwidth]{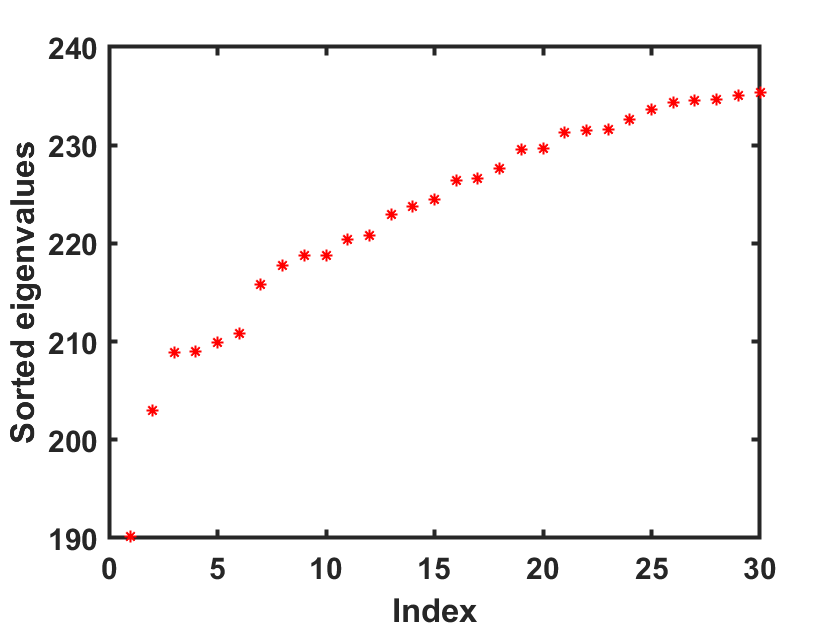}
    \caption{
    First thirty ordered eigenvalues for the Laplacian of an unweighted $K$NN graph from the experiment in 
    Figure~\ref{fig:2NNnoise} based on data points from $[0,1]^{15}$. 
    }
    \label{fig:gaptwoNNnoise}
\end{figure}

With this addition of noise,
Algorithm~\ref{alg:spectwoNN}
gives a reasonable estimate of 
the intrinsic dimension $d=15$, as shown in Figure~\ref{fig:2NNnoise}.
In Figure~\ref{fig:gaptwoNNnoise} we see, however, that the spectrum of the Laplacian 
does not give compelling evidence.
Intuitively, the twoNN construction should be less sensitive to the presence of a small percentage of missing or spurious edges, since these perturbations will not affect the majority of the first or second pairwise distances.

\section{A Test on Real Data}\label{sec:mnist}

\begin{figure}[htp]
    \centering
    \includegraphics[width=1\textwidth]{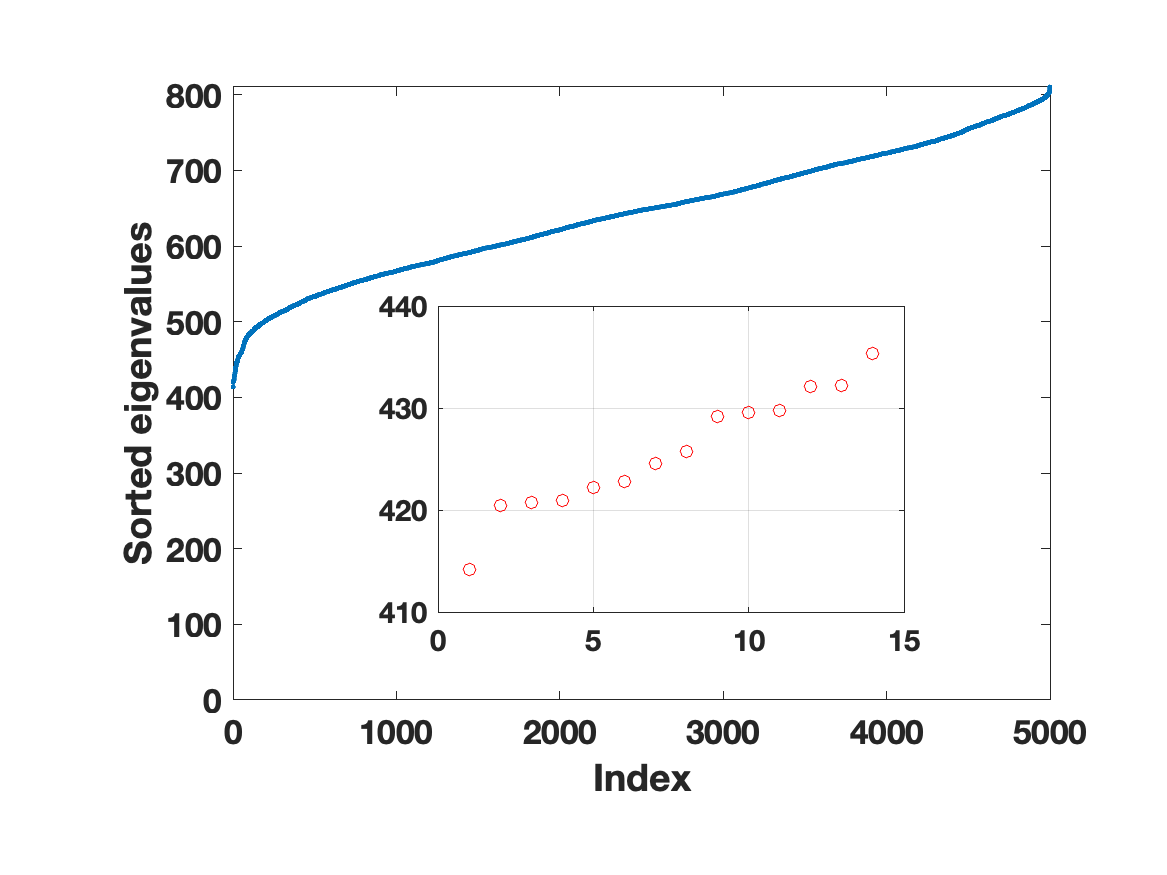}
    \caption{Ordered nonzero eigenvalues of the Laplacian based on inverse pairwise Euclidean distances between $N = 5,000$ handwritten digits from 0 to 9.
    Interior plot zooms in on the smallest eigenvalues.}
    \label{fig:mnistlapa}
\end{figure}

\begin{figure}[htp]
    \centering
    \includegraphics[width=1\textwidth]{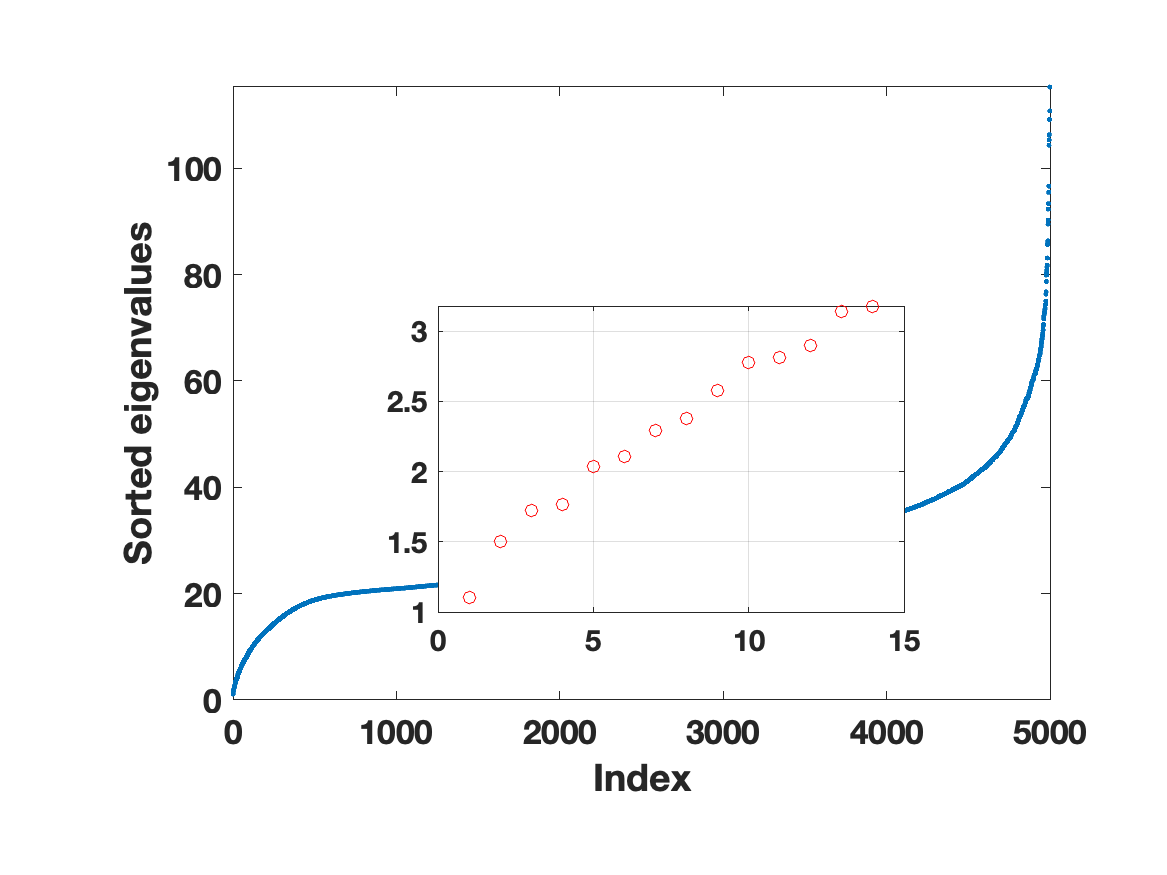}
    \caption{Ordered nonzero eigenvalues of the Laplacian based on $K = 20$ nearest neighbour graph from $N = 5,000$ handwritten digits from 0 to 9.
    Interior plot zooms in on the smallest eigenvalues.}
    \label{fig:mnistlapb}
\end{figure}

We finish with a test on MNIST image data \cite{lcb-digits_old}. 
Here each data point in $\RR^{784}$ represents the $28 \times 28$ greyscale pixel values of a handwritten digit, from 0 to 9. 
We uploaded  
5,000 images using 
\texttt{digitTrain4DArrayData}
in MATLAB \cite{MATLAB22}, where each image has been arbitrarily rotated.
Figure~\ref{fig:mnistlapa} shows the Laplacian spectrum for the similarity matrix based on the reciprocal of Euclidean distance.
We then constructed the binarized $K = 20$ nearest neighbour graph. Figure~\ref{fig:mnistlapb} shows the corresponding spectrum.
We see that neither spectrum plot shows a definitive spectral gap, and the two plots are not consistent.

\begin{figure}[htp]
    \centering
    \includegraphics[width=1\textwidth]{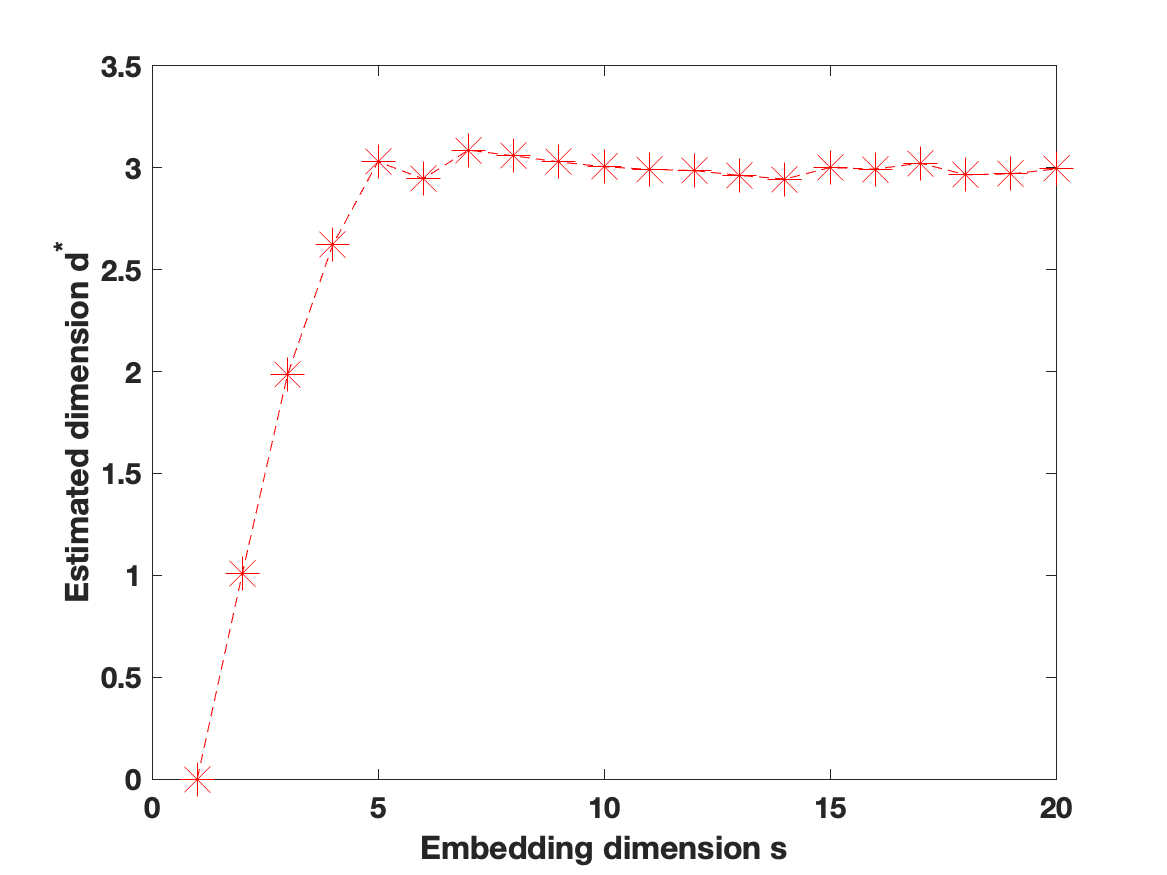}
    \caption{ Results from Algorithm~\ref{alg:spectwoNN} with dimension $s$ varying from  
    $5$ to $30$ on the horizontal axis. Here we used the same nearest neighbour graph as in Figure~\ref{fig:mnistlapb}.
    }
    \label{fig:mnist2nn}
\end{figure}

Applying twoNN to the Euclidean distance data gave a dimension estimate of $d^\star = 3.37$.
For the $K = 20$ nearest neighbour version of the data, Figure~\ref{fig:mnist2nn} shows the results from
Algorithm~\ref{alg:spectwoNN} as the embedding dimension $s$ varies from  
    $5$ to $20$. We see that on this binarized data, it is possible to use twoNN in a way that gives an estimate that is consistent with the 
   result on the original pairwise distance data.

   We emphasize that the aim of this test was not to produce a definitive result for the underlying dimension of MNIST. This issue has been tackled 
   in a number of works, including \cite{HA05,PZAGG21}, with answers that depend on the way that the concept of dimension is introduced.
   We aimed to test instead whether Algorithm~\ref{alg:spectwoNN} allows twoNN to remain consistent under K nearest neighbour binarization.
   We note however, that the MNIST data can be embedded into 3 dimensional space in a visually pleasing manner\footnote{https://colah.github.io/posts/2014-10-Visualizing-MNIST/}.

\section{Conclusions}\label{sec:conc}

The twoNN algorithm in \cite{Facco17} 
gives a computationally efficient way to 
estimate the dimension of a data cloud, assuming the points are samples from 
a continuous manifold. 
The algorithm requires only first and second nearest neighbour distances; for a 
weighted (undirected) network this information is immediately available and we found that the algorithm performed well
on examples where a ground truth is available and where information from the Laplacian spectrum was not useful. 
For unweighted networks, where edges are either present or absent, the algorithm can no longer be applied directly. However, we showed that 
consistent estimates of 
the dimension can be recovered by spectrally embedding into successively higher dimensional Euclidean space and applying twoNN
at each stage. 
We also found that this approach was more robust to noise 
than the direct use of the Laplacian spectrum,
and more consistent under $K$ nearest neighbour binarization.

Overall, these results highlight and extend the usefulness of twoNN in the context of network analysis.

\section*{Acknowledgements} 
  DJH was supported by the Engineering and Physical Sciences Research Council under grants EP/P020720/1 and EP/V015605/1.
  We thank Clive Bowman for comments on this work.

\bibliographystyle{siam}
\bibliography{dimrefs}

\end{document}